\documentclass{article}

\usepackage[margin=1in]{geometry}
\usepackage{graphicx}
\usepackage{braket}
\usepackage{amsmath}
\usepackage{amsfonts}
\usepackage{mathtools}
\usepackage{amsthm}
\usepackage{xcolor}
\usepackage{placeins}
\usepackage{tabularx}
\usepackage{booktabs}
\usepackage{multicol}
\usepackage{comment}
\usepackage{titlesec}
\usepackage{authblk}
\usepackage{tikz}
\usepackage{etoolbox}

\usetikzlibrary{
    arrows.meta,
    calc,
    decorations.pathmorphing,
    decorations.pathreplacing
}

\titleformat{\section}
  {\normalfont\bfseries\centering}
  {\Roman{section}.}
  {0.75em}
  {\MakeUppercase}

\titleformat{\subsection}
  {\normalfont\bfseries\centering}
  {\Alph{subsection}.}
  {0.75em}
  {}

\newtheorem{theorem}{Theorem}[section]

\newtheorem{corollary}[theorem]{Corollary}
\newtheorem{proposition}[theorem]{Proposition}

\theoremstyle{definition}

\theoremstyle{remark}
\newtheorem{remark}[theorem]{Remark}

\usepackage[
    backend=biber,
    style=numeric-comp,
    sorting=none,
    giveninits=true,
    maxbibnames=9,
    doi=false,
    isbn=false,
    url=false
]{biblatex}

\AtBeginBibliography{%
    \footnotesize
    \setlength{\bibitemsep}{0pt}
    \setlength{\bibparsep}{0pt}
    \linespread{0.90}\selectfont
}

\usepackage[hidelinks]{hyperref}

\title{
    Gravitationally Mediated Entanglement
    in Interferometers with Repeated Splitting
}

\author[1]{Riccardo Zampieri\thanks{%
    Present address of studies:
    Department of Applied Mathematics and Theoretical Physics,
    Centre for Mathematical Sciences,
    University of Cambridge,
    Wilberforce Road,
    Cambridge CB3 0WA,
    United Kingdom.
}}

\author[1,2]{Sougato Bose}

\affil[1]{
    Department of Physics and Astronomy,
    University College London,
    Gower Street,
    London WC1E 6BT,
    United Kingdom}
\affil[2]{David Potter Institute for Quantum Information and Spacetime,
University College London, Gower Street, WC1E 6BT London, United Kingdom
}

\date{}

\begin{document}

\maketitle

\begin{abstract}
Proposals to probe the quantum nature of gravity through entanglement
between spatially separated masses have attracted considerable interest.
In such schemes, spatial superpositions are prepared through controlled
splitting operations that produce several interferometric branches.
Here we propose and analyse protocols involving repeated splitting
stages, each of which doubles the number of occupied paths.

We show that the density matrix after $n$ splitting stages is related
to that of the $(n-1)$-splitting protocol through local parity
dephasing. Two of the resulting sectors reproduce the preceding state,
while the other two acquire additional gravitational phases. Removing
the locally generated parts gives an exact lower bound on the change
in negativity, expressed in terms of the preceding state's negative
partial-transpose eigenspace. For two splitting stages, this leads to
an explicit condition on three gravitational phases under which the
additional split cannot reduce negativity. We also show that further
splitting cannot shorten the lifetime of the generated entanglement.

If the local internal system preserves the full branch label, the
positional entanglement is retained upon recombination. We describe
an implementation using a reusable two-level system and local memories,
and compare it with a readout without a memory register. For the latter, we derive an exact condition for when an additional splitting stage improves the readout, based on a $2\times 2$ Schur complement. Numerical simulations show an advantage in positional
negativity over the rates considered, with an improvement surviving the
restricted readout at all decoherence rates considered at the final readout time.
\end{abstract}
\newpage

\section{Introduction}

The relationship between quantum theory and gravity remains one of the
central open problems in modern physics. The possibility of probing quantum
aspects of gravity in the laboratory was already seriously discussed by
Feynman and others at the 1957 Chapel Hill conference~\cite{dewitt1957}, and
the quantum-mechanical sensitivity of matter to classical gravitational potentials was
subsequently demonstrated in the Colella--Overhauser--Werner
experiment~\cite{colella1975}. At the same time, more direct probes of the
gravitational field at the quantum level, such as the detection of individual
gravitons, are usually thought to be far beyond present experimental reach because of the
extreme weakness of the gravitational interaction~\cite{dyson2013}, although there are some recent interesting developments \cite{tobar2024detecting}.

This has motivated a different strategy: rather than attempting to observe
quantum gravity directly, one may ask whether gravity can mediate
distinctively quantum effects between material systems. In these
proposals, spatially separated masses are prepared in spatial superposition
and allowed to interact only through their mutual gravitational
field~\cite{bose2016matter,bose2017,marletto2017}. If the masses become entangled, then, as explained in \cite{bose2017,marshman2020},
under the usual quantum-information reasoning that entanglement cannot be
generated by local operations and classical communication
(LOCC)~\cite{horodecki2009}, the mediator cannot be purely classical. In fact, under the assumption of no direct action at a distance non-local interactions, it is not possible to have an operator valued interaction between two masses, which is necessary to produce entanglement, unless the mediator of this interaction is quantum \cite{bose2022}. Further justifications have also been given, eg, \cite{christodoulou2019,christodoulou2023local,galley2022}. Consistent alternatives to quantum gravity, such as 
postquantum classical gravity~\cite{oppenheim2023} make the need for such experiments more acute. The interpretation of gravity-mediated entanglement also depends on the assumptions made about locality and the interaction~\cite{hallreginatto2018,martinmartinez2023}. Tests
of the quantum measurement postulate for gravity \cite{hanif2024testing}, and weak-value effects of
gravitational quantum superpositions~\cite{bosevaidman2026repulsive}, which use similar technology, have also been proposed. 

Matter-wave interferometry provides a natural setting in which to pursue
these questions. Spin-dependent Stern--Gerlach splitting offers coherent
control over spatial superpositions while retaining access to internal degrees of
freedom that can later be used as entanglement
witnesses~\cite{machluf2013,margalit2021,scala2013,marshman2020}. These ideas
have been extended theoretically to mesoscopic masses and long interaction
times~\cite{wan2016,carney2019}. Recent work has analysed gravitationally
induced entanglement in interacting Stern--Gerlach interferometers at the
density-matrix level, including open dynamics, mass-independent leading-order
entangling phases, and recombination limitations~\cite{braccini2026mass}.
Other work has addressed practical noise and implementation constraints,
including acceleration and tilt noise, magnetic noise, inertial torsion
noise, phonon-induced contrast loss, dipole--dipole decoherence, air
scattering, and electromagnetic screening~\cite{moorthy2026random,
moorthy2025magnetic,wu2025torsion,xiang2024phonon,fragolino2024dipole,
schut2024air,schut2023screening,schut2024micrometer}.

Several works have explored configurations beyond the standard two-path
setting. Entanglement generation through central-potential interactions has
been analysed in mesoscopic systems~\cite{qvarfort2020central}. It has been
shown that qudit encodings can outperform two-qubit superpositions in the
presence of strong decoherence~\cite{tilly2021}, that a third mass can
improve resilience~\cite{schut2022}, and that suitable multi-particle
arrangements can favour entanglement generation~\cite{li2023}. A relativistic
variant based on superpositions of rotational energies has also been
proposed~\cite{higgins2024}. These results indicate that the geometry of the
spatial superposition is a relevant design choice.

The question we address is whether the {\em number} of splitting stages
can itself increase the entanglement generated by the gravitational
interaction. We consider binary internal-state-dependent splitting,
where each stage doubles the number of spatial paths. A concrete
implementation could reuse one active two-level system while storing
earlier branch choices in a local memory register, building on
coherent electron--nuclear spin control in diamond
\cite{fuchs2011memory,taminiau2014register,bradley2019register}.
We use the parity structure of the resulting state to compare one
splitting stage with the next. After removing phases that act locally,
we obtain a general bound on the change in negativity, an explicit
condition for two stages, and a result on the persistence of
entanglement. We then consider recombination. A full internal register
retains the positional entanglement, while a two-level readout without
memory combines different branches and may lose part of it. We reduce the question of whether its conditioned negativity improves
to a $2\times2$ matrix test. Finally, we simulate both protocols along
finite-duration trajectories, comparing their negativities and the
analytical conditions as the decoherence rate varies. The numerical
parameters are chosen for an illustrative comparison with a gravitational
interaction larger than the estimated Casimir--Polder background.

\newpage

\section{Mathematical Framework}
\label{sec:framework}

We consider two subsystems, labelled $A$ and $B$, each carrying a local
internal system together with the positional modes of an interferometer. For
$X\in\{A,B\}$, let $\mathcal H_X^{\mathrm{int}}$ denote the internal Hilbert
space. After $k$ splitting stages, define
\begin{equation}
\label{eq:hilbert}
\mathcal H_{X,\mathrm{pos}}^{(k)}
:=
\mathrm{span}\{|m\rangle_X\}_{m=0}^{2^k-1},
\qquad
\mathcal H_X^{(k)}
:=
\mathcal H_X^{\mathrm{int}}
\otimes
\mathcal H_{X,\mathrm{pos}}^{(k)} .
\end{equation}
Thus
$\mathcal H_{X,\mathrm{pos}}^{(0)}
=\mathrm{span}\{|0\rangle_X\}
=\mathrm{span}\{|C\rangle_X\}$,
and each splitting stage doubles the dimension of the occupied positional
space.

Each occupied positional branch has a definite locally associated internal
state. We write it as $|\chi_m^{(k)}\rangle_X$ after $k$ stages, so the
corresponding full local state is
$|\chi_m^{(k)}\rangle_X\otimes|m\rangle_X$. Because distinct positional
modes are orthogonal, identifying these occupied joint states with the
abstract branch labels $|m\rangle_X$ is a local isometry. We use this
identification throughout: internal labels are suppressed when they play no
role in the spatial dynamics, but they are not traced out.

The integer $n$ denotes the maximal number of coherent splitting stages. The
joint state after the $n$th stage belongs to
\begin{equation}
\label{eq:total_hilbert}
\mathcal H^{(n)}
=
\mathcal H_A^{(n)}\otimes\mathcal H_B^{(n)}.
\end{equation}
We use the product branch basis
\begin{equation}
\label{eq:basis}
|i,j\rangle:=|i\rangle_A\otimes|j\rangle_B,
\qquad
i,j\in\{0,\dots,2^n-1\},
\end{equation}
and denote the corresponding branch-space matrix elements of $\rho_n$ by
\begin{equation}
\label{eq:matrix_elements}
(\rho_n)_{ij,kl}:=\langle i,j|\rho_n|k,l\rangle.
\end{equation}

\subsection{Forward Splitting}
\label{sec:forward_split}

We distinguish a reusable two-level degree of freedom $q$, which generates the
state-dependent spatial splitting, from a local memory register. For a
fully branch-preserving $n$-stage implementation one may take
\begin{equation}
\label{eq:internal_register}
\mathcal H_X^{\mathrm{int}}
=
\mathcal H_{q,X}
\otimes
\bigotimes_{r=1}^{n-1}\mathcal H_{M_r,X},
\end{equation}
where the memory factors store branch choices from earlier stages. The same
active two-level system can then be returned to a fiducial state and reused
for each subsequent split.

Suppose that before stage $k$ the previous branch choices are encoded in a
memory state $|\mathbf b(m)\rangle_M$, where $m\in\{0,\ldots,2^{k-1}-1\}$
is the parent-branch index, and the active system is in $|0\rangle_q$. We
allow the local rotation phase to depend on $m$ and write
\begin{equation}
\label{eq:active_rotation}
|\mathbf b(m)\rangle_M|0\rangle_q
\longmapsto
\frac{|\mathbf b(m)\rangle_M}{\sqrt2}
\left(
|0\rangle_q+e^{i\theta_{k,m}}|1\rangle_q
\right).
\end{equation}
The state-dependent splitting then acts on the active system and the
centre-of-mass mode as
\begin{equation}
\label{eq:controlled_split}
|b\rangle_q|m\rangle_X
\longmapsto
|b\rangle_q|2m+b\rangle_X,
\qquad b\in\{0,1\}.
\end{equation}
For $k<n$, the newly generated branch bit can be transferred coherently to a
fresh memory qubit $M_k$, initially in $|0\rangle_{M_k}$, through the local
storage operation
\begin{equation}
\label{eq:memory_transfer}
W_k:\qquad
|b\rangle_q|0\rangle_{M_k}
\longmapsto
|0\rangle_q|b\rangle_{M_k},
\qquad b\in\{0,1\}.
\end{equation}
The active system is thereby returned to $|0\rangle_q$ and may be reused at
the next stage. Coherent transfer between an NV electronic spin and a nuclear
spin memory, together with control of multi-qubit electron--nuclear registers
in diamond, has been demonstrated experimentally
\cite{fuchs2011memory,taminiau2014register,bradley2019register}. We idealise
this storage as a fast local operation; any finite transfer time may be
included in the inter-stage delay while the positional branches are held
fixed.

After the final stage, if $i=(b_1\cdots b_n)_2$, the complete branch label may
therefore be retained in the local internal state
\begin{equation}
\label{eq:branch_register_state}
|\chi_i^{(n)}\rangle_X
=
|b_1\rangle_{M_1}\cdots|b_{n-1}\rangle_{M_{n-1}}|b_n\rangle_q,
\end{equation}
which gives $2^n$ mutually orthogonal internal states. A uniform rotation at
stage $k$ corresponds to $\theta_{k,m}=\theta_k$. We nevertheless retain the
more general $\theta_{k,m}$ in Eq.~\eqref{eq:active_rotation}; it allows local
parent-branch-dependent rephasings and includes the restricted two-level
implementation considered below.

Suppressing the internal labels, the induced action on the occupied branch
subspace is exactly
\begin{equation}
\label{eq:split_action}
V_k^{(X)}|m\rangle_X
=
\frac{1}{\sqrt{2}}
\left(
|2m\rangle_X
+
e^{i\theta_{k,m}}|2m+1\rangle_X
\right),
\qquad
0\le m\le 2^{k-1}-1.
\end{equation}
A Stern--Gerlach operation provides one physical realization of this
state-dependent spatial splitting
\cite{machluf2013,margalit2021,scala2013,marshman2020}. The proofs below use
only the induced action in Eq.~\eqref{eq:split_action} and do not depend on a
particular realization of the storage operation.

\begin{figure*}[h]
    \centering
    \resizebox{0.96\textwidth}{!}{%
    \begin{tikzpicture}[
        >=Latex,
        traj/.style={line width=0.9pt},
        branchzero/.style={traj},
        branchone/.style={traj},
        guide/.style={densely dotted, gray, line width=0.55pt},
        grav/.style={
            decorate,
            decoration={
                snake,
                amplitude=1.3pt,
                segment length=5pt
            },
            line width=0.8pt
        },
        smalllab/.style={
            font=\scriptsize,
            inner sep=1.2pt
        },
        measure/.style={
            draw,
            circle,
            line width=0.8pt,
            inner sep=2.5pt,
            font=\small
        }
    ]


    \draw[-{Latex[length=3mm]}, line width=0.8pt]
        (-6.35,5.15) -- (-6.35,-2.18);

    \node[font=\small, anchor=east] at (-6.45,5.08) {$t=0$};
    \node[font=\small, anchor=east] at (-6.45,-2.18) {$t$};

    \draw[<->, line width=0.7pt]
        (-4,5.25) -- (4,5.25)
        node[midway, fill=white, inner sep=2pt] {$d_0$};

    \node[font=\large] at (-4,5.65) {$A$};
    \node[font=\large] at ( 4,5.65) {$B$};

    \draw[guide, rounded corners=2pt]
        (-5.55,4.68) rectangle (-2.45,-1.65);

    \draw[guide, rounded corners=2pt]
        ( 2.45,4.68) rectangle ( 5.55,-1.65);


    \foreach \X/\name in {-4/A,4/B} {

        \node[smalllab, above] at (\X,4.80)
            {$|C\rangle_{\name}$};

        \draw[traj]
            (\X,4.75) -- (\X,4.55);

        \draw[branchzero]
            (\X,4.55) -- ({\X-0.75},3.85);

        \draw[branchone]
            (\X,4.55) -- ({\X+0.75},3.85);

        \draw[branchzero]
            ({\X-0.75},3.85) -- ({\X-0.75},3.35);

        \draw[branchone]
            ({\X+0.75},3.85) -- ({\X+0.75},3.35);

        \draw[branchzero]
            ({\X-0.75},3.35) -- ({\X-1.25},2.55);

        \draw[branchone]
            ({\X-0.75},3.35) -- ({\X-0.35},2.55);

        \draw[branchzero]
            ({\X+0.75},3.35) -- ({\X+0.35},2.55);

        \draw[branchone]
            ({\X+0.75},3.35) -- ({\X+1.25},2.55);

        \draw[branchzero]
            ({\X-1.25},2.55) -- ({\X-1.25},0.20);

        \draw[branchone]
            ({\X-0.35},2.55) -- ({\X-0.35},0.20);

        \draw[branchzero]
            ({\X+0.35},2.55) -- ({\X+0.35},0.20);

        \draw[branchone]
            ({\X+1.25},2.55) -- ({\X+1.25},0.20);

        \node[smalllab, fill=white] at (\X,2.45)
            {$|x_0\rangle,\ldots,|x_3\rangle$};

        \draw[branchzero]
            ({\X-1.25},0.20) -- ({\X-0.75},-0.55);

        \draw[branchone]
            ({\X-0.35},0.20) -- ({\X-0.75},-0.55);

        \draw[branchzero]
            ({\X+0.35},0.20) -- ({\X+0.75},-0.55);

        \draw[branchone]
            ({\X+1.25},0.20) -- ({\X+0.75},-0.55);

        \draw[traj]
            ({\X-0.75},-0.55) -- (\X,-1.35);

        \draw[traj]
            ({\X+0.75},-0.55) -- (\X,-1.35);

        \draw[traj]
            (\X,-1.35) -- (\X,-1.80);

        \node[measure] at (\X,-2.02) {$M_{\name}$};
    }


    \draw[grav]
        (-2.75,1.25) -- (2.75,1.25);

    \node[font=\small, fill=white, inner sep=2pt]
        at (0,1.80)
        {$E_{ij}=\dfrac{\kappa}{d_{ij}}$};

    \node[font=\scriptsize, fill=white, inner sep=1.5pt]
        at (0,1.22)
        {all branch pairs $(i,j)$};


    \draw[
        decorate,
        decoration={brace,mirror,amplitude=5pt},
        line width=0.7pt
    ]
        (-4.65,-2.47) -- (4.65,-2.47)
        node[
            midway,
            below=8pt,
            font=\small
        ]
        {internal-state correlation measurement};

    \end{tikzpicture}%
    }
    \caption{
    Schematic of the repeated-splitting interferometric protocol, shown for two splitting stages ($n=2$). 
    Two masses, $A$ and $B$, initially separated by $d_0$, begin in localised centre-of-mass states $|C\rangle_A$ and $|C\rangle_B$. 
    At each forward stage, a reusable two-level degree of freedom controls a binary spatial split; between stages, earlier branch choices may be transferred coherently to a local memory register, so that after two stages each subsystem occupies four positional modes while retaining the complete branch label. 
    During the interaction interval, each branch pair $(i,j)$ accumulates a gravitationally induced phase determined by the interaction energy $E_{ij}=\kappa/d_{ij}$. 
    The motional branches are subsequently recombined for readout. The branch geometry is common to both readout cases discussed after the positional-mode analysis.
    }
    \label{fig:multisplit_setup}
\end{figure*}
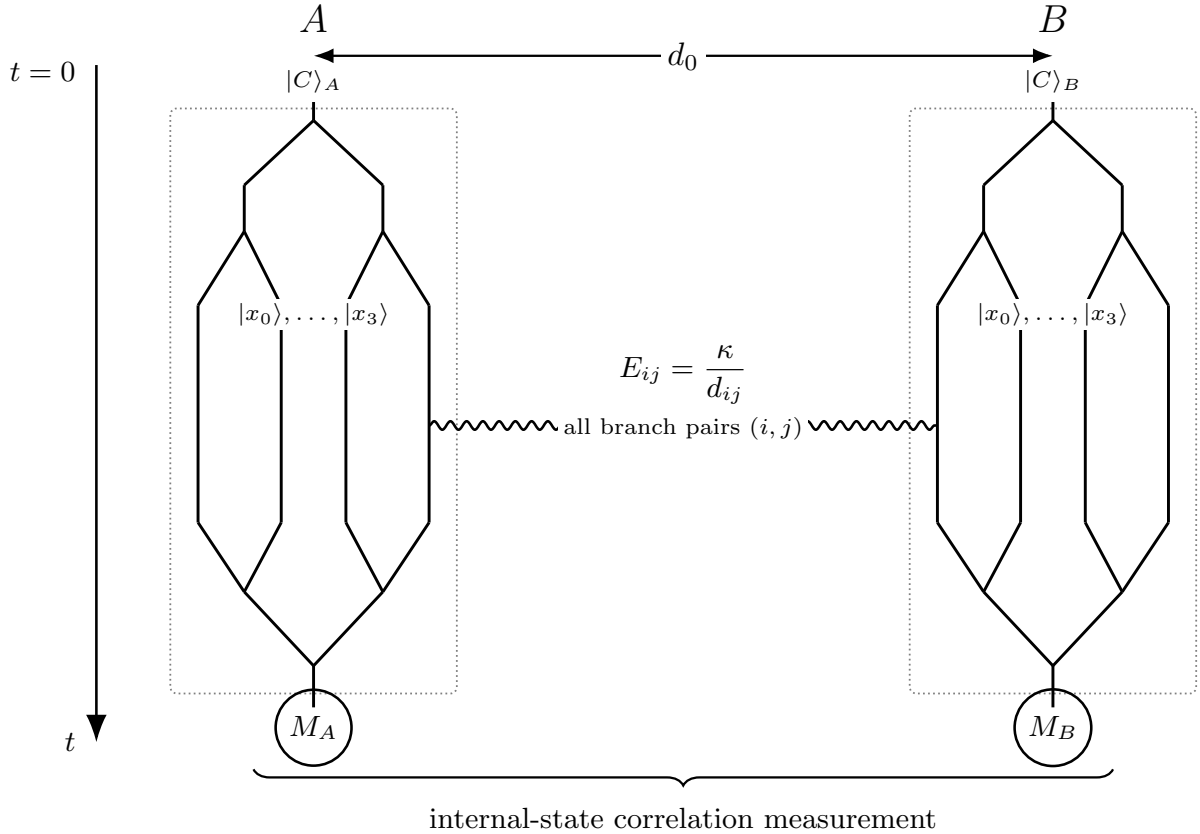

\subsection{Time Evolution}
\label{sec:time_evolution}

We denote by $\tau_n^{(1)}$ the time at which the $n$th splitting stage
begins, with $\tau_1^{(1)}=0$, and by $\tau_n^{(2)}$ the time at which branches originating from the same parent have reached their maximum separation after the $n$th split. After this time, branches are left to evolve under the interaction described below with no further changes in separation until either the $(n+1)$th split or recombination occurs. The interaction responsible for phase
accumulation is diagonal in the product position basis and proportional
to the inverse inter-system distance $d_{ij}(t)$:
\begin{equation}
\label{eq:hint}
H_{\mathrm{int}}(t)
=
\sum_{i,j=0}^{2^n-1}
E_{ij}(t)\,|i,j\rangle\langle i,j|
\otimes I_{AB}^{\mathrm{int}}
=
\sum_{i,j=0}^{2^n-1}
\frac{\kappa}{d_{ij}(t)}
\,|i,j\rangle\langle i,j|
\otimes I_{AB}^{\mathrm{int}},
\end{equation}
for some constant $\kappa$.

Environmental decoherence suppresses spatial coherences between distinct
branches. We model this by a time-local Lindbladian
\cite{joos1985,hornberger2008,romero2011a} acting independently on the
two subsystems:
\begin{equation}
\label{eq:lindblad}
\mathcal{L}_{\mathrm{dec}}[\rho]
=
-\sum_{i,k}
\Gamma_A(x_i-x_k,t)\,
|i\rangle\langle i|_A\,\rho\,|k\rangle\langle k|_A
-\sum_{j,l}
\Gamma_B(x_j-x_l,t)\,
|j\rangle\langle j|_B\,\rho\,|l\rangle\langle l|_B.
\end{equation}
This term leaves populations invariant and damps only off-diagonal
coherences in the positional basis. The full evolution is therefore
\begin{equation}
\label{eq:master}
\frac{d\rho}{dt}
=
-\frac{i}{\hbar}[H_{\mathrm{int}}(t),\rho]
+\mathcal{L}_{\mathrm{dec}}[\rho],
\end{equation}
which yields the elementwise equation
\begin{equation}
\label{eq:elementwise}
\frac{d}{dt}\rho_{ij,kl}
=
-\frac{i}{\hbar}\bigl(E_{ij}(t)-E_{kl}(t)\bigr)\rho_{ij,kl}
-
\bigl[
\Gamma_A(x_i-x_k,t)+\Gamma_B(x_j-x_l,t)
\bigr]\rho_{ij,kl}.
\end{equation}
Integrating gives
\begin{align}
\label{eq:formal_solution}
\rho_{ij,kl}(t)
&=
\rho_{ij,kl}(0)
\exp\!\left(
-\int_0^t
\frac{i}{\hbar}\bigl(E_{ij}(s)-E_{kl}(s)\bigr)\,ds
\right)
\nonumber\\
&\qquad\times
\exp\!\left(
-\int_0^t
\bigl[
\Gamma_A(x_i-x_k,s)+\Gamma_B(x_j-x_l,s)
\bigr]\,ds
\right).
\end{align}

For the localisation rate we use the standard phenomenological
form~\cite{romero2011a,schut2024air}
\begin{equation}
\label{eq:gamma}
\Gamma_X(\Delta x)
=
\begin{cases}
\Lambda_X\,\Delta x^2,
& |\Delta x|\ll 2a_X, \\[6pt]
\gamma_X
\left(
1-\exp\!\left[-\dfrac{\Delta x^2}{4a_X^2}\right]
\right),
& |\Delta x|\sim 2a_X, \\[8pt]
\gamma_X,
& |\Delta x|\gg 2a_X,
\end{cases}
\end{equation}
where $\Lambda_X=\gamma_X/(4a_X^2)$, $a_X$ is the localisation length
scale, and $\gamma_X$ is the maximal decoherence rate. For identical
subsystems in a common environment we assume that
\begin{equation}
\label{eq:symmetry}
a_A=a_B\equiv a,
\qquad
\gamma_A=\gamma_B\equiv\gamma.
\end{equation}

\begin{proposition}[Free evolution]
\label{prop:free_evolution}
After the $m$th splitting stage, the system evolves freely during the
interval $\tau_m^{(2)}\le t<\tau_{m+1}^{(1)}$. In this regime the
positions of all branches are fixed, so the pairwise separations
$d_{ij}$ and intra-subsystem separations are constant in time. Consequently the interaction energies
$E_{ij}=\kappa/d_{ij}$ and the local decoherence rates
$\Gamma_X(x_i-x_k)$ are also time-independent. It follows from
Eq.~\eqref{eq:master} that, for $\tau_m^{(2)}\le t<\tau_{m+1}^{(1)}$,
the density-matrix elements evolve as
\[
\rho_{ij,kl}(t)
=
\rho_{ij,kl}(\tau_m^{(2)})
\exp\!\left[
-\frac{i}{\hbar}
\left(
\frac{\kappa}{d_{ij}}-\frac{\kappa}{d_{kl}}
\right)
(t-\tau_m^{(2)})
\right]
\exp\!\left[
-\bigl(\Gamma_A(x_i-x_k)+\Gamma_B(x_j-x_l)\bigr)
(t-\tau_m^{(2)})
\right].
\]
Thus, once a splitting stage has ended, the subsequent dynamics reduce
to coherent phase accumulation determined by the interaction energies
together with exponential suppression of off-diagonal coherences at
rates fixed by the branch separations reached at the end of the split.
\end{proposition}

\section{Entanglement in the Positional Modes}
\label{sec:ent_dyn}

The analysis in this section concerns the occupied spatial branch structure.
Locally attached internal labels are carried along through the isometric
identification introduced in Section~\ref{sec:framework}; they are not traced
out. The resulting $A|B$ negativity is independent of the dimension of the
local internal system and of the local splitting phases $\theta_{k,m}$.

\subsection{Parity Dephasing}

To analyse the entanglement structure of the multi-splitting protocol
it is convenient to group positional modes according to their parity.
After $n$ coherent splitting stages the positional Hilbert space of
each subsystem is
$\mathcal H_{X,\mathrm{pos}}^{(n)}
= \mathrm{span}\{|m\rangle_X\}_{m=0}^{N-1}$
with $N = 2^n$. We partition the positional basis into even and odd
mode indices,
\begin{equation}
\label{eq:EO_partition}
E := \{\,m \in \{0,\dots,N-1\} : m\ \text{even}\,\}, \qquad
O := \{\,m \in \{0,\dots,N-1\} : m\ \text{odd}\,\},
\end{equation}
inducing the orthogonal decomposition
$\mathcal H_{X,\mathrm{pos}}^{(n)}
= \mathcal H_{X,E}\oplus\mathcal H_{X,O}$.
Define the corresponding projectors
\begin{equation}
\label{eq:parity_projectors}
P^{(X)}_{E} := \sum_{m \in E} |m\rangle_X\!\langle m|, \qquad
P^{(X)}_{O} := \sum_{m \in O} |m\rangle_X\!\langle m|,
\end{equation}
extended trivially over the local internal system as
$\Pi^{(X)}_{E} := I^{\mathrm{int}}_X \otimes P^{(X)}_{E}$ and
$\Pi^{(X)}_{O} := I^{\mathrm{int}}_X \otimes P^{(X)}_{O}$, where
$I_X^{\mathrm{int}}$ is the identity on whichever internal states are present.

The (non-conditioned) LOCC channel corresponding to locally measuring
even/odd positional parity on each subsystem and discarding the
classical outcomes is
\begin{equation}
\label{eq:locc_parity_channel}
\mathcal{M}(\rho)
= \sum_{x,y \in \{E,O\}}
(\Pi^{(A)}_{x} \otimes \Pi^{(B)}_{y})\,\rho\,
(\Pi^{(A)}_{x} \otimes \Pi^{(B)}_{y}),
\end{equation}
with Kraus operators $K_{xy} := \Pi^{(A)}_{x} \otimes \Pi^{(B)}_{y}$
satisfying $\sum_{x,y} K_{xy}^\dagger K_{xy} = I_{AB}$, so
$\mathcal{M}$ is completely positive and trace-preserving. In the
product position basis its action on matrix elements is
\begin{equation}
\label{eq:parity_selection_rule}
\bigl[\mathcal{M}(\rho)\bigr]_{ij,kl}
=
\begin{cases}
\rho_{ij,kl}, & \mathrm{par}(i)=\mathrm{par}(k)\ \text{and}\
\mathrm{par}(j)=\mathrm{par}(l),\\[4pt]
0, & \text{otherwise},
\end{cases}
\end{equation}
where $\mathrm{par}(m) = 0$ for $m$ even and $\mathrm{par}(m) = 1$
for $m$ odd. Thus $\mathcal{M}$ implements a local parity dephasing
of the joint state with respect to
$\mathcal{H}_{A,E} \oplus \mathcal{H}_{A,O}$ and
$\mathcal{H}_{B,E} \oplus \mathcal{H}_{B,O}$.

We quantify bipartite entanglement in the positional modes using the
negativity \cite{vidalwerner2002,peres1996,PlenioVirmani2007} with respect to the
partition $A|B$,
\begin{equation}
\label{eq:negativity_def}
\mathcal{N}(\rho) := \frac{\|\rho^{T_A}\|_* - 1}{2},
\end{equation}
where $T_A$ denotes partial transpose on subsystem $A$ and
$\|\cdot\|_*$ the trace norm. Since $\mathcal{M}$ is LOCC, negativity
is monotonic under its action,
\begin{equation}
\label{eq:neg_monotonicity}
\mathcal{N}\!\bigl(\mathcal{M}(\rho)\bigr) \le \mathcal{N}(\rho).
\end{equation}

Define the four (generally unnormalised) post-measurement states
\begin{equation}
\label{eq:rho_xy_def}
\rho_{xy} :=
(\Pi^{(A)}_x \otimes \Pi^{(B)}_y)\,\rho\,
(\Pi^{(A)}_x \otimes \Pi^{(B)}_y),
\qquad x,y \in \{E,O\},
\end{equation}
so that $\mathcal{M}(\rho) = \rho_{EE} + \rho_{EO} + \rho_{OE} +
\rho_{OO}$. Since the projectors $\Pi^{(A)}_x$ are diagonal in the
position basis they are invariant under transpose, so
$(\mathcal{M}(\rho))^{T_A}$ is block-diagonal with respect to the
parity sectors. Additivity of the trace norm for block-diagonal
operators \cite{Bhatia1997} then gives
\begin{equation}
\label{eq:trace_norm_additivity}
\bigl\|(\mathcal{M}(\rho))^{T_A}\bigr\|_*
= \sum_{x,y} \bigl\|(\rho_{xy})^{T_A}\bigr\|_*,
\end{equation}
and setting $p_{xy} := \mathrm{Tr}(\rho_{xy})$ and
$\tilde\rho_{xy} := \rho_{xy}/p_{xy}$ for $p_{xy} > 0$,
\begin{equation}
\label{eq:negativity_decomposition}
\mathcal{N}\!\bigl(\mathcal{M}(\rho)\bigr)
= \sum_{x,y \in \{E,O\}} p_{xy}\,\mathcal{N}(\tilde\rho_{xy}).
\end{equation}
The LOCC channel thus removes all coherences between different parity
sectors while preserving the entanglement internal to each block.
We now consider the effect of $\mathcal{M}$ on $\rho_n$. For any
branch index $m$ we write $\tilde{m} := \lfloor m/2 \rfloor$ for its
parent index in the preceding splitting stage; thus $\tilde\imath,
\tilde\jmath, \tilde{k}, \tilde{l}$ denote the parents of $i, j, k,
l$ respectively.

Before analysing each block, we record a key geometric observation
that underlies all four cases. Under the $n$th splitting, all branches
of a given parity are accelerated in the same direction: even branches
move one way and odd branches the other. Consequently, within any
single subsystem, two branches of the same parity have a separation
that is not only constant throughout the splitting stage, but
identical to the separation between their respective parent branches
in the $(n-1)$-splitting protocol. This is because same-parity
branches receive identical displacements, so no additional relative
separation is generated between them by the splitting pulse. This
means that for every block, the intra-subsystem separations
$x_i - x_k$ (subsystem $A$) and $x_j - x_l$ (subsystem $B$) satisfy
$x_i - x_k = x_{\tilde\imath} - x_{\tilde{k}}$ and
$x_j - x_l = x_{\tilde\jmath} - x_{\tilde{l}}$,
and remain so during subsequent free evolution by
Proposition~\ref{prop:free_evolution}. The decoherence rates
$\Gamma_A(x_i - x_k)$ and $\Gamma_B(x_j - x_l)$ are therefore
time-independent and equal to those of the parent protocol in all
four blocks. The inter-system distances $d_{ij}(t)$, by contrast,
depend on the relative parity of the subsystem $A$ and subsystem $B$
branches: same-parity pairs (EE and OO) also receive identical
displacements and their inter-system separation remains constant and
equal to $d_{\tilde\imath\tilde\jmath}$, while opposite-parity pairs
(EO and OE) move in opposite directions and their separation acquires
a time-dependent shift $\pm x(t)$ relative to $d_{\tilde\imath\tilde\jmath}$.
This distinction between the diagonal and off-diagonal blocks is the
source of the phase correction $\delta\Phi^{EO}_{ij,kl}$ derived below.

\paragraph*{The EE block.}

We suppress the internal degree of freedom momentarily. The even--even
block has matrix elements
\begin{equation}
\label{eq:EE_block_elements}
(\rho_{EE})_{ij,kl} =
\delta_{i \in E}\,\delta_{k \in E}\,\delta_{j \in E}\,\delta_{l \in E}
\,\rho_{ij,kl}.
\end{equation}
In this block both subsystems contribute even branches, which move in
the same direction during the $n$th splitting. The inter-system
distances $d_{ij}$ are therefore constant throughout the splitting
stage and subsequent free evolution, and the interaction energies
$E_{ij} = \kappa/d_{ij}$ are time-independent. By
Proposition~\ref{prop:free_evolution}, the density-matrix elements
for $t \ge \tau_n^{(2)}$ evolve as
\begin{align}
\label{eq:EE_time_evolution}
(\rho_{EE}(t))_{ij,kl} &= (\rho_{EE}(\tau_n^{(2)}))_{ij,kl}
\exp\!\left[
-\frac{i}{\hbar}\!\left(
\frac{\kappa}{d_{ij}}-\frac{\kappa}{d_{kl}}
\right)(t-\tau_n^{(2)})
\right] \nonumber \\
&\quad\times
\exp\!\left[
-\bigl(\Gamma_A(x_i-x_k)+\Gamma_B(x_j-x_l)\bigr)
(t-\tau_n^{(2)})
\right].
\end{align}
From the construction in Section~\ref{sec:forward_split}, every even
child produced by the binary splitting inherits its parent's
amplitude multiplied by $1/\sqrt{2}$, so each density-matrix element
indexed by four even states acquires an overall factor of $1/4$:
\begin{equation}
\label{eq:EE_initial_condition}
(\rho_{EE}(\tau_n^{(2)}))_{ij,kl} =
\frac{1}{4}
(\rho_{n-1}(\tau_n^{(2)}))_{\tilde\imath\tilde\jmath,\tilde{k}\tilde{l}}.
\end{equation}
Applying Proposition~\ref{prop:free_evolution} to $\rho_{n-1}$,
which evolves freely after $\tau_{n-1}^{(2)}$,
\begin{align}
\label{eq:EE_backprop}
(\rho_{EE}(\tau_n^{(2)}))_{ij,kl} &=
\frac{1}{4}
(\rho_{n-1}(\tau_{n-1}^{(2)}))_{\tilde\imath\tilde\jmath,\tilde{k}\tilde{l}}
\exp\!\left[
-\frac{i}{\hbar}\!\left(
\frac{\kappa}{d_{\tilde\imath\tilde\jmath}}
-\frac{\kappa}{d_{\tilde{k}\tilde{l}}}
\right)(\tau_n^{(2)}-\tau_{n-1}^{(2)})
\right] \nonumber \\
&\quad\times
\exp\!\left[
-\bigl(\Gamma_A(x_{\tilde\imath}-x_{\tilde{k}})
+\Gamma_B(x_{\tilde\jmath}-x_{\tilde{l}})\bigr)
(\tau_n^{(2)}-\tau_{n-1}^{(2)})
\right].
\end{align}
Combining with Eq.~\eqref{eq:EE_initial_condition} and
Eq.~\eqref{eq:EE_time_evolution} yields
\begin{equation}
\label{eq:EE_recursion}
(\rho_{EE}(t))_{ij,kl} =
\frac{1}{4}(\rho_{n-1}(t))_{\tilde\imath\tilde\jmath,\tilde{k}\tilde{l}},
\end{equation}
so that, up to the relabelling $m \mapsto \tilde{m}$,
\begin{equation}
\label{eq:EE_block_final}
\rho_{EE} = \tfrac{1}{4}\,\rho_{n-1}.
\end{equation}
The internal labels introduced by the $n$th split are local to each
subsystem. Relabelling them to match the parent protocol is therefore a local
operation and does not change the $A|B$ negativity.

\paragraph*{The OO block.}

The odd--odd block follows identically. Both subsystems contribute odd
branches, which also move in the same direction during the $n$th
splitting, so the inter-system distances are again constant and
Proposition~\ref{prop:free_evolution} applies without modification.

The only difference arises at the splitting step. An odd child of parent
$m$ acquires the factor $e^{i\theta_{n,m}}/\sqrt{2}$ at the $n$th
stage. It follows that
\begin{equation}
\label{eq:OO_recursion}
(\rho_{OO}(t))_{ij,kl}
=
\frac{
e^{i(
\theta_{n,\tilde\imath}
+\theta_{n,\tilde\jmath}
-\theta_{n,\tilde{k}}
-\theta_{n,\tilde{l}}
)}
}{4}
(\rho_{n-1}(t))_{\tilde\imath\tilde\jmath,\tilde{k}\tilde{l}}.
\end{equation}
The additional factor separates into local diagonal rephasings on subsystems
$A$ and $B$. Thus $\rho_{OO}=\tfrac14\rho_{n-1}$ up to relabelling and
local unitaries, which do not change the negativity.

\paragraph*{The EO block.}
In the even--odd sector ($i,k\in E$, $j,l\in O$), the two masses receive
opposite displacements, so their separations increase by the same amount
$x_n(s)$ for every parent pair:
\begin{equation}
\label{eq:EO_distance_shift}
d_{ij}(s)=d_{\tilde\imath\tilde\jmath}+x_n(s),\qquad
d_{kl}(s)=d_{\tilde k\tilde l}+x_n(s).
\end{equation}
The intra-subsystem branch separations, and hence their decoherence rates,
remain those of the parent protocol.

For a Stern--Gerlach realisation, let
\begin{equation}
\label{eq:SG_acceleration}
\alpha_{\mathrm{SG}}=\frac{\mu_{\mathrm{eff}}\mathcal G}{m},
\qquad \mathcal G=\partial_xB.
\end{equation}
With $s=t-\tau_n^{(1)}$, we take acceleration for $0\le s\le T_\alpha$,
free drift for $T_\alpha\le s\le T_\alpha+T_d$, and symmetric
deceleration for $T_\alpha+T_d\le s\le2T_\alpha+T_d$.
The resulting relative displacement is
\begin{equation}
\label{eq:x_piecewise}
x_n(\tau_n^{(1)}+s)=
\begin{cases}
\alpha_{\mathrm{SG}}s^2, &0\le s\le T_\alpha,\\[3pt]
\alpha_{\mathrm{SG}}T_\alpha^2+2\alpha_{\mathrm{SG}}T_\alpha(s-T_\alpha),
&T_\alpha\le s\le T_\alpha+T_d,\\[3pt]
\begin{aligned}
&\alpha_{\mathrm{SG}}T_\alpha^2+2\alpha_{\mathrm{SG}}T_\alpha T_d\\
&+2\alpha_{\mathrm{SG}}T_\alpha(s-T_\alpha-T_d)\\
&-\alpha_{\mathrm{SG}}(s-T_\alpha-T_d)^2,
\end{aligned}
&T_\alpha+T_d\le s\le2T_\alpha+T_d,
\end{cases}
\end{equation}
with $x_n(u)=0$ before the stage and, after its completion,
\begin{equation}
\label{eq:x_final}
x_n(u)=x_n^{\mathrm{peak}}
=2\alpha_{\mathrm{SG}}T_\alpha(T_\alpha+T_d).
\end{equation}
In subsequent integrals, $x_n(u)$ is the displacement at absolute
time $u$, with the pulse trajectory given above at
$s=u-\tau_n^{(1)}$.

For any parent pair $(p,q)$, define the additional gravitational phase
\begin{equation}
\label{eq:phase_pair_exact}
F_\pm(p,q;t)=\frac{\kappa}{\hbar}
\int_{\tau_n^{(1)}}^t
\left[\frac1{d_{pq}\pm x_n(u)}-\frac1{d_{pq}}\right]du.
\end{equation}
The plus sign describes EO. In particular, the correction to the relative
phase of the matrix element is
\begin{equation}
\label{eq:dPhiEO}
\delta\Phi^{EO}_{ij,kl}(t)=
F_+(\tilde\imath,\tilde\jmath;t)-F_+(\tilde k,\tilde l;t).
\end{equation}
The exact EO block is therefore
\begin{equation}
\label{eq:EO}
(\rho_{EO})_{ij,kl}
=\frac{e^{i(\theta_{n,\tilde\jmath}-\theta_{n,\tilde l})}}4
(\rho_{n-1})_{\tilde\imath\tilde\jmath,\tilde k\tilde l}
 e^{-i\delta\Phi^{EO}_{ij,kl}}.
\end{equation}
Thus EO is a rescaled copy of the parent state acted upon by a diagonal
gravity-induced unitary, apart from a local rotation phase.

\paragraph*{The OE block.}
For odd--even branches the separation decreases:
\begin{equation}
\label{eq:OE_distance_shift}
d_{ij}(s)=d_{\tilde\imath\tilde\jmath}-x_n(s),\qquad
d_{kl}(s)=d_{\tilde k\tilde l}-x_n(s).
\end{equation}
The same reasoning gives
\begin{equation}
\label{eq:dPhiOE}
\delta\Phi^{OE}_{ij,kl}(t)=
F_-(\tilde\imath,\tilde\jmath;t)-F_-(\tilde k,\tilde l;t)
\end{equation}
and
\begin{equation}
\label{eq:OE}
(\rho_{OE})_{ij,kl}
=\frac{e^{i(\theta_{n,\tilde\imath}-\theta_{n,\tilde k})}}4
(\rho_{n-1})_{\tilde\imath\tilde\jmath,\tilde k\tilde l}
 e^{-i\delta\Phi^{OE}_{ij,kl}}.
\end{equation}
The same-parity decoherence rates equal those of the parent also in this
sector. No expansion in $x_n/d_{pq}$ has been used. Although
$F_-=-F_+$ to first order in the new displacement, their exact
magnitudes generally differ at higher orders.

\subsection{Removing Local Phases}
\label{subsec:local_phases}
Let us look more closely at the additional phases $F_\pm$ in the EO
and OE blocks. Some of their contributions act on one mass only.
We may remove these by local unitaries without changing the
entanglement. For example, suppose the interaction is quadratic in
the inter-mass distance,
$V_2(D)=c_0+c_1(D-d_0)+\frac12c_2(D-d_0)^2$.
Writing the parent branch positions relative to their centres as
$a_p,b_q$, so that $d_{pq}=d_0+b_q-a_p$, a relative displacement
$z$ changes the potential by
\begin{equation}
\label{eq:quadratic_local}
\begin{aligned}
V_2(d_{pq}+z)-V_2(d_{pq})
&=c_1z+c_2z(b_q-a_p)+\tfrac12c_2z^2\\
&=\underbrace{-c_2za_p}_{A\text{-only}}
+\underbrace{c_2zb_q}_{B\text{-only}}
+\underbrace{c_1z+\tfrac12c_2z^2}_{\text{common}}.
\end{aligned}
\end{equation}
The same separation holds after integrating over time. Thus the
additional phases of a quadratic potential can be removed entirely
by local rephasings. For a general potential, a part depending on
both branch choices may remain.

To identify that part, regard $F_\pm(p,q)$ in
Eq.~\eqref{eq:phase_pair_exact} as a function of the parent positions,
$F_\pm(a_p,b_q)$. A phase of the form $A(a_p)+B(b_q)$ is removable:
its first term acts on $A$ alone and its second on $B$ alone.
Such terms cancel in the mixed difference
\begin{equation}
\label{eq:mixed_difference}
F_\pm(a,b)-F_\pm(a,b')-F_\pm(a',b)+F_\pm(a',b').
\end{equation}
The converse also holds. If every mixed difference is zero, fix
$a',b'$ and rearrange to write $F_\pm(a,b)$ as a function of $a$
plus a function of $b$. This gives an exact test for which phases
cannot be removed locally, for any number of branches.

There is some freedom in choosing the local phases to subtract.
We use a symmetric choice based on the outermost \emph{parent}
branches: $a_-=\min_p a_p$, $a_+=\max_p a_p$, and likewise
$b_-,b_+$ for mass $B$. Average the mixed differences obtained
by taking each of these four corners as the reference:
\begin{equation}
\label{eq:R_rectangle}
\begin{split}
R_\pm(a,b)=\frac14\sum_{i,j\in\{-,+\}}\big[
&F_\pm(a,b)-F_\pm(a,b_j)\\
&-F_\pm(a_i,b)+F_\pm(a_i,b_j)\big].
\end{split}
\end{equation}
Collecting terms gives the decomposition
\begin{equation}
\label{eq:phase_centring}
F_\pm(a,b)=A_\pm(a)+B_\pm(b)+R_\pm(a,b),
\end{equation}
where
\begin{align}
\label{eq:local_phase_choices}
A_\pm(a)&=\tfrac12[F_\pm(a,b_-)+F_\pm(a,b_+)],\\
B_\pm(b)&=\tfrac12[F_\pm(a_-,b)+F_\pm(a_+,b)]
-\tfrac14\!\sum_{i,j\in\{-,+\}}F_\pm(a_i,b_j).
\end{align}
By construction, $A_\pm$ depends only on mass $A$, $B_\pm$
only on mass $B$, and $R_\pm$ contains the remaining mixed phase.

These phase functions define diagonal operators on the parent
branch basis:
\begin{equation}
\label{eq:phase_operators}
\begin{aligned}
A_\pm&=\sum_p A_\pm(a_p)|p\rangle\langle p|_A\otimes I_B,\\
B_\pm&=I_A\otimes\sum_q B_\pm(b_q)|q\rangle\langle q|_B,\\
R_\pm&=\sum_{p,q}R_\pm(a_p,b_q)|pq\rangle\langle pq|.
\end{aligned}
\end{equation}
The operator $F_\pm$ is their sum. Since all three are diagonal
in the same basis,
$[A_\pm,B_\pm]=[A_\pm,R_\pm]=[B_\pm,R_\pm]=0$.
We may therefore factorise the exponential:
\begin{equation}
\label{eq:phase_factorization}
e^{-iF_\pm}=e^{-iA_\pm}e^{-iB_\pm}e^{-iR_\pm}.
\end{equation}
The first two factors act separately on $A$ and $B$.
After measuring the final parities, we may undo them in the EO
and OE sectors, together with the local splitting phases in
Eqs.~\eqref{eq:EO} and \eqref{eq:OE}. The two normalised
mixed-sector states are then
\begin{equation}
\label{eq:phase_unitaries}
\sigma_\pm=U_\pm\sigma U_\pm^\dagger,
\qquad U_\pm=e^{-iR_\pm},
\qquad \sigma=\rho_{n-1}(t).
\end{equation}
The EE and OO sectors are already local copies of $\sigma$.
Each parity outcome has probability $1/4$. Since parity
measurement and the conditional rephasings are LOCC, the
additivity of negativity across the four blocks, shown in
Eq.~\eqref{eq:negativity_decomposition}, now gives
\begin{equation}
\label{eq:exact_sector_bound}
\begin{aligned}
\mathcal N(\rho_n)
&\ge\mathcal N(\mathcal M(\rho_n))\\
&=\tfrac12\mathcal N(\sigma)
+\tfrac14\mathcal N(\sigma_+)
+\tfrac14\mathcal N(\sigma_-).
\end{aligned}
\end{equation}

\subsection{One-Step Variation of Entanglement}
\label{subsec:monotonicity}
We can now compare the two protocols directly. Subtracting
$\mathcal N(\sigma)$ from Eq.~\eqref{eq:exact_sector_bound},
\begin{equation}
\label{eq:sector_difference_bound}
\mathcal N(\rho_n)-\mathcal N(\sigma)
\ge\tfrac14\sum_{\pm}
[\mathcal N(\sigma_\pm)-\mathcal N(\sigma)].
\end{equation}
To bound these two differences, recall the variational form of
negativity~\cite{vidalwerner2002,PlenioVirmani2007,Bhatia1997}:
\begin{equation}
\label{eq:neg_variational}
\mathcal N(\tau)=\max_{0\le P\le I}
[-\operatorname{Tr}(P\tau^{T_A})].
\end{equation}
Indeed, if $\tau^{T_A}=\sum_j\lambda_j|j\rangle\langle j|$,
then $-\operatorname{Tr}(P\tau^{T_A})
=-\sum_j\lambda_j\langle j|P|j\rangle$ is largest when
$P$ projects onto the eigenvectors with negative eigenvalues.
Their absolute eigenvalues sum to $\mathcal N(\tau)$.

\begin{theorem}[Successive negativity bound]
\label{thm:monotone}
Let $\sigma=\rho_{n-1}(t)$ and let $\sigma_\pm$ be the
corrected sector states in Eq.~\eqref{eq:phase_unitaries}.
Let $P_-$ project onto the negative eigenspace of
$\sigma^{T_A}$, with $P_-=0$ if that eigenspace is empty. Then
\begin{equation}
\label{eq:W_definition}
\begin{aligned}
\mathcal N(\rho_n)-\mathcal N(\rho_{n-1})&\ge-\frac12\operatorname{Tr}\!\left[
P_-\left(\frac{\sigma_++\sigma_-}{2}-\sigma\right)^{T_A}
\right] := W_n.
\end{aligned}
\end{equation}
If $W_n\ge0$ then another splitting stage does not reduce the negativity.
\end{theorem}

\begin{proof}
The choice $P_-$ gives the parent's negativity exactly:
$\mathcal N(\sigma)=-\operatorname{Tr}(P_-\sigma^{T_A})$.
For each corrected sector it is still an allowed choice in
Eq.~\eqref{eq:neg_variational}, whether or not it is optimal.
Thus
\begin{equation}
\label{eq:W_variational_step}
\begin{aligned}
\mathcal N(\sigma_\pm)-\mathcal N(\sigma)
&\ge-\operatorname{Tr}(P_-\sigma_\pm^{T_A})
+\operatorname{Tr}(P_-\sigma^{T_A})\\
&=-\operatorname{Tr}\!\left[
P_-(\sigma_\pm-\sigma)^{T_A}\right].
\end{aligned}
\end{equation}
Summing these two inequalities with weight $1/4$ as in
Eq.~\eqref{eq:sector_difference_bound}, and combining their
traces, gives
\[
-\frac14\sum_{\pm}\operatorname{Tr}\!\left[
P_-(\sigma_\pm-\sigma)^{T_A}\right]
=-\frac12\operatorname{Tr}\!\left[
P_-\left(\frac{\sigma_++\sigma_-}{2}-\sigma\right)^{T_A}
\right].
\]
This proves the bound.
\end{proof}

The subtraction in Eq.~\eqref{eq:sector_difference_bound}
explains our choice of $P_-$. It gives the parent's negativity
\emph{exactly}, leaving only the corrected sectors to estimate.
A projector chosen for a corrected sector might improve that estimate,
but would generally no longer evaluate the parent term exactly.
When $\sigma^{T_A}$ has no zero eigenvalues, $P_-$ also gives
the first-order change in each sector's negativity for sufficiently
small corrections. A negative $W_n$ does not imply decreasing
entanglement: the corrected states may have different negative
eigenspaces. If $P_-=0$, the parent negativity is zero and
non-decrease is automatic.

\paragraph*{Small-displacement expansion.}
For a small additional split, we can evaluate the witness
without computing the full exponentials. Suppose
$|x_n(u)|$ is small compared with every occupied inter-mass
distance, at fixed comparison time. Equation~\eqref{eq:phase_pair_exact}
then gives
\begin{equation}
\label{eq:F_small_expansion}
F_\pm(p,q)=\frac\kappa\hbar
\int_{\tau_n^{(1)}}^t\!du\left[
\mp\frac{x_n(u)}{d_{pq}^2}
+\frac{x_n(u)^2}{d_{pq}^3}
\mp\frac{x_n(u)^3}{d_{pq}^4}+\cdots\right].
\end{equation}
The subtraction defining $R_\pm$ is linear, so the terms
odd in $x_n$ have opposite signs in the two sectors, while the
even terms have the same sign. Accordingly, define
\begin{equation}
\label{eq:RsRa}
R_{\mathrm{s}}=\tfrac12(R_++R_-)=O((x_n^{\mathrm{peak}})^2),
\quad
R_{\mathrm{a}}=\tfrac12(R_--R_+)=O(x_n^{\mathrm{peak}}).
\end{equation}
Both are diagonal and therefore commute. As
$R_\pm=R_{\mathrm{s}}\mp R_{\mathrm{a}}$, the average
appearing in the theorem can be written as
\begin{equation}
\label{eq:symmetric_conjugation}
\begin{split}
\frac{\sigma_++\sigma_-}{2}
=e^{-iR_{\mathrm{s}}}\bigg[
&\frac12e^{iR_{\mathrm{a}}}\sigma e^{-iR_{\mathrm{a}}}\\
+&\frac12e^{-iR_{\mathrm{a}}}\sigma e^{iR_{\mathrm{a}}}
\bigg]e^{iR_{\mathrm{s}}}.
\end{split}
\end{equation}
Expanding each conjugation term by term,
\begin{equation}
\label{eq:unitary_commutator_series}
\begin{aligned}
e^{-iR}\sigma e^{iR}
&=(I-iR-\tfrac12R^2+\cdots)\sigma
  (I+iR-\tfrac12R^2+\cdots)\\
&=\sigma-i[R,\sigma]
-\tfrac12[R,[R,\sigma]]
+\tfrac i6[R,[R,[R,\sigma]]]+\cdots.
\end{aligned}
\end{equation}
The odd powers of $R_{\mathrm{a}}$ cancel when the two terms
inside the brackets in Eq.~\eqref{eq:symmetric_conjugation}
are averaged:
\begin{equation}
\label{eq:odd_cancel}
\begin{aligned}
&\tfrac12\big(e^{iR_{\mathrm{a}}}\sigma e^{-iR_{\mathrm{a}}}
+e^{-iR_{\mathrm{a}}}\sigma e^{iR_{\mathrm{a}}}\big)\\
&\qquad=\sigma-\tfrac12[R_{\mathrm{a}},[R_{\mathrm{a}},\sigma]]
+O((x_n^{\mathrm{peak}})^4).
\end{aligned}
\end{equation}
Now apply the remaining conjugation by $e^{-iR_{\mathrm{s}}}$.
Because $R_{\mathrm{s}}$ is second order, its first
commutator contributes at second order, while its double
commutator and its action on the $R_{\mathrm{a}}$ double
commutator are both fourth order. Hence
\begin{equation}
\label{eq:sector_expansion}
\begin{aligned}
\frac{\sigma_++\sigma_-}{2}
&=\sigma-i[R_{\mathrm{s}},\sigma]
-\tfrac12[R_{\mathrm{a}},[R_{\mathrm{a}},\sigma]]\\
&\qquad+O((x_n^{\mathrm{peak}})^4).
\end{aligned}
\end{equation}
Subtracting $\sigma$ in Theorem~\ref{thm:monotone}
and using the linearity of the partial transpose gives
\begin{equation}
\label{eq:W_quadratic}
\begin{aligned}
W_n={}&\ \tfrac i2\operatorname{Tr}\!\left[
P_-[R_{\mathrm{s}},\sigma]^{T_A}\right]\\
&+\tfrac14\operatorname{Tr}\!\left[
P_-[R_{\mathrm{a}},[R_{\mathrm{a}},\sigma]]^{T_A}\right]
+O((x_n^{\mathrm{peak}})^4).
\end{aligned}
\end{equation}
Which may be used as an approximation to the exact witness.

\begin{corollary}[Persistence of negativity]
\label{cor:persistence}
If $\mathcal N(\rho_{n-1}(t))>0$, then
$\mathcal N(\rho_n(t))>0$.
\end{corollary}
\begin{proof}
The two same-parity sectors alone give
$\mathcal N(\rho_n)\ge\frac12\mathcal N(\rho_{n-1})$
from Eq.~\eqref{eq:exact_sector_bound}. Thus if the $n-1$ split negativity is nonzero, so is the $n$ split negativity.
Equivalently, the partial transpose of the EE (or OO) block of $\rho_n^{T_A}$ is a principal
submatrix, locally equivalent to $\rho_{n-1}^{T_A}/4$.
Cauchy's interlacing theorem~\cite{Bhatia1997} then shows directly
that a negative eigenvalue of the parent gives a negative eigenvalue
of $\rho_n^{T_A}$.
\end{proof}
Thus an extra splitting stage cannot shorten the interval over
which negativity is nonzero.

\subsection{Analytical Condition for Two Splitting Stages}
\label{subsec:two_split_criterion}

We now apply the general result to $n=2$. The parent state
$\sigma=\rho_1$ has two branches per mass, so the residual
phases $R_\pm$ and the negative eigenspace of $\sigma^{T_A}$
can both be found explicitly.

Let $x_1(u)$ and $x_2(u)$ be the relative separations generated
by the two splitting stages, following the finite-duration
trajectories of Eq.~\eqref{eq:x_piecewise}. By the time the
second split begins, the first has reached
$x_1^{\mathrm{peak}}$. The four parent branch-pair distances,
ordered as $(00,01,10,11)$, are therefore
\[
d_0,\qquad d_0+x_1^{\mathrm{peak}},\qquad
d_0-x_1^{\mathrm{peak}},\qquad d_0.
\]
For each pair, the additional gravitational phase is
$F_\pm(p,q)$ from Eq.~\eqref{eq:phase_pair_exact}, with these
distances substituted. Since there are only two parent branches
per mass, the symmetric subtraction in
Eq.~\eqref{eq:R_rectangle} gives
\begin{equation}
\label{eq:R_two_split_F}
\begin{aligned}
R_\pm={}&\frac{
F_\pm(0,0)-F_\pm(0,1)-F_\pm(1,0)+F_\pm(1,1)
}{4}\\
&\times
\begin{pmatrix}
1&0&0&0\\
0&-1&0&0\\
0&0&-1&0\\
0&0&0&1
\end{pmatrix}.
\end{aligned}
\end{equation}

We can express this combination more conveniently in terms
of the gravitational phases accumulated by the two-path
parent and the corrected sectors. In each case the
nonlocal phase is the sum of the phases of the opposite-parity
branch pairs minus those of the equal-parity pairs.
We denote these three phases by
\begin{equation}
\label{eq:two_qubit_phases}
\begin{split}
\Phi_1(t)=\frac{\kappa}{\hbar}\int_0^t\!du\,\bigg[
&\frac1{d_0-x_1(u)}
+\frac1{d_0+x_1(u)}
-\frac2{d_0}\bigg],\\
\Phi_\pm(t)=\frac{\kappa}{\hbar}\int_0^t\!du\,\bigg[
&\frac1{d_0\pm x_2(u)-x_1(u)}
+\frac1{d_0\pm x_2(u)+x_1(u)}\\
&\hspace{26mm}-\frac2{d_0\pm x_2(u)}\bigg].
\end{split}
\end{equation}
The first is the gravitational phase of the single-split
protocol. The other two are the corresponding phases in EO
and OE, where the second split shifts the inter-mass
distance by $\pm x_2(u)$. Here $x_k(u)=0$ before stage $k$
begins and remains at $x_k^{\mathrm{peak}}$ after its
completion, so these integrals include the full splitting
dynamics.

Comparing Eq.~\eqref{eq:two_qubit_phases} with the definition
of $F_\pm$, the additional phase combination in
Eq.~\eqref{eq:R_two_split_F} is
\[
\begin{aligned}
&F_\pm(0,0)-F_\pm(0,1)
-F_\pm(1,0)+F_\pm(1,1)\\
&\hspace{30mm}=-(\Phi_\pm-\Phi_1).
\end{aligned}
\]
Indeed, $\Phi_\pm$ and $\Phi_1$ agree before the second
split, while their difference afterwards is precisely the
negative of the four $F_\pm$ terms above. We therefore obtain
\begin{equation}
\label{eq:R_two_split}
R_\pm=-\frac{\Phi_\pm-\Phi_1}{4}
\begin{pmatrix}
1&0&0&0\\
0&-1&0&0\\
0&0&-1&0\\
0&0&0&1
\end{pmatrix}.
\end{equation}
After local rephasing, the parent's gravitational amplitudes
are proportional to
$e^{i\Phi_1/4},e^{-i\Phi_1/4},
e^{-i\Phi_1/4},e^{i\Phi_1/4}$.
Equation~\eqref{eq:R_two_split} shows that multiplication
by $e^{-iR_\pm}$ simply replaces $\Phi_1$ by $\Phi_\pm$.

We can now obtain the three density matrices directly from
the evolution in Section~\ref{sec:time_evolution}. The local
path coherence is suppressed by
\begin{equation}
\label{eq:eta_one}
\eta(t)=\exp\left[
-\int_0^t\Gamma(x_1(u))\,du
\right].
\end{equation}
A coherence between states differing in one local branch
index acquires one factor of $\eta$, while one differing in
both indices acquires $\eta^2$, since the two dephasing
processes act independently. These factors are the same in
all three states: within each parity sector the separation
between parent branches remains $x_1(u)$.

In the ordered basis $(|00\rangle,|01\rangle,|10\rangle,
|11\rangle)$, and after the local rephasings described above,
we may therefore write
\begin{equation}
\label{eq:two_qubit_matrix}
\sigma(\Phi)=\frac14
\begin{pmatrix}
1&\eta e^{i\Phi/2}&\eta e^{i\Phi/2}&\eta^2\\
\eta e^{-i\Phi/2}&1&\eta^2&\eta e^{-i\Phi/2}\\
\eta e^{-i\Phi/2}&\eta^2&1&\eta e^{-i\Phi/2}\\
\eta^2&\eta e^{i\Phi/2}&\eta e^{i\Phi/2}&1
\end{pmatrix},
\end{equation}
where $\sigma=\sigma(\Phi_1)$ and
$\sigma_\pm=\sigma(\Phi_\pm)$. The equal-parity branch pairs
carry gravitational amplitudes $e^{i\Phi/4}$, while the
opposite-parity pairs carry $e^{-i\Phi/4}$. Their relative
phases give the factors $e^{\pm i\Phi/2}$ in the matrix.

Taking the partial transpose gives the eigenvalues
\begin{equation}
\label{eq:two_qubit_eigenvalues}
\frac{1+\eta^2\pm2\eta\cos(\Phi/2)}4,
\qquad
\frac{1-\eta^2\pm2\eta\sin(\Phi/2)}4.
\end{equation}
The first pair is always nonnegative. The parent therefore
has a negative eigenvalue precisely when
\begin{equation}
\label{eq:parent_npt_condition}
2\eta\left|\sin\frac{\Phi_1}{2}\right|>1-\eta^2.
\end{equation}
If this condition is not satisfied,
$\mathcal N(\rho_1)=0$, and the second split cannot
reduce its negativity.

Suppose instead that the parent has a negative eigenvalue.
For $0<\Phi_1<2\pi$, this eigenvalue and its normalised
eigenvector are
\begin{equation}
\label{eq:negative_eigen_two_qubit}
\lambda_-=
\frac{1-\eta^2-2\eta\sin(\Phi_1/2)}4,
\qquad
|v_-\rangle=
\frac{|00\rangle+i|01\rangle-i|10\rangle-|11\rangle}{2}.
\end{equation}
In this local phase convention the eigenvector is independent
of $\Phi_1$ and $\eta$. Direct substitution into the partial
transpose of Eq.~\eqref{eq:two_qubit_matrix} gives
\[
\langle v_-|\sigma(\Phi)^{T_A}|v_-\rangle
=\frac{1-\eta^2-2\eta\sin(\Phi/2)}4
\]
for any $\Phi$.

Using $P_-=|v_-\rangle\langle v_-|$ in
Theorem~\ref{thm:monotone}, we obtain
\begin{equation}
\label{eq:W_2_exact}
\begin{aligned}
W_2=-\frac14\big[&
\langle v_-|\sigma(\Phi_+)^{T_A}|v_-\rangle\\
&+\langle v_-|\sigma(\Phi_-)^{T_A}|v_-\rangle\\
&-2\langle v_-|\sigma(\Phi_1)^{T_A}|v_-\rangle
\big]\\
=\frac{\eta}{8}\bigg[
&\sin\frac{\Phi_+}{2}
+\sin\frac{\Phi_-}{2}
-2\sin\frac{\Phi_1}{2}
\bigg].
\end{aligned}
\end{equation}
The terms $1-\eta^2$ cancel between the two corrected
sectors and twice the parent. Thus, for an entangled parent
with $0<\Phi_1<2\pi$, a sufficient condition for the second
split not to reduce negativity is
\begin{equation}
\label{eq:n2_exact_condition}
\sin\frac{\Phi_+}{2}
+\sin\frac{\Phi_-}{2}
\ge 2\sin\frac{\Phi_1}{2}.
\end{equation}

If the parent is entangled outside this phase interval, the
negative eigenvector changes when $\sin(\Phi_1/2)$ changes
sign. Replacing $i$ by $-i$ in
Eq.~\eqref{eq:negative_eigen_two_qubit} gives the corresponding
projector and hence
\begin{equation}
\label{eq:W_2_extended}
W_2=\frac{\eta}{8}
\operatorname{sgn}\!\left(\sin\frac{\Phi_1}{2}\right)
\left[
\sin\frac{\Phi_+}{2}
+\sin\frac{\Phi_-}{2}
-2\sin\frac{\Phi_1}{2}
\right].
\end{equation}
Thus, when $\sin(\Phi_1/2)<0$, the inequality
\eqref{eq:n2_exact_condition} is reversed.
At $\sin(\Phi_1/2)=0$ the parent has zero negativity,
as already covered above.

Finally, for small gravitational phases,
Eq.~\eqref{eq:n2_exact_condition} reduces at leading order to
$\Phi_++\Phi_-\ge2\Phi_1$, consistent with the convexity
of the Newtonian inverse-distance interaction.

\begin{figure}[!htbp]
\centering
\includegraphics[width=\linewidth]{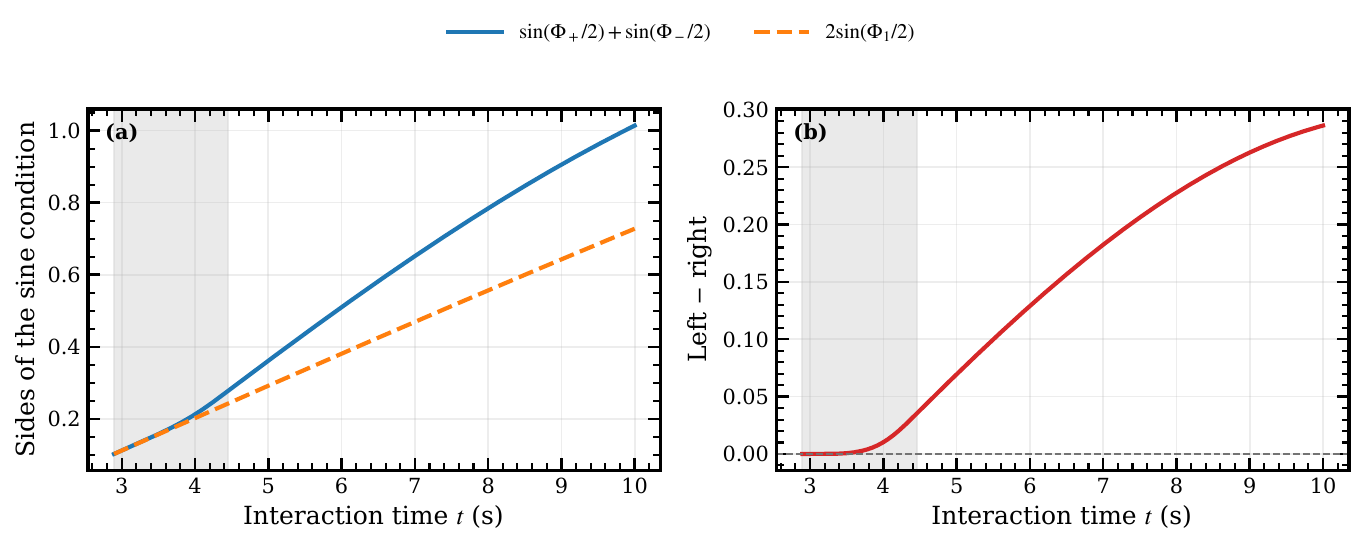}
\caption{The sufficient condition in Eq.~\eqref{eq:n2_exact_condition}
for the parameters of Table~\ref{tab:params} in Section~\ref{sec:numerical}.
\textbf{(a)} The two sides of the inequality; \textbf{(b)} their
difference, which remains nonnegative over the interval shown.
The shaded region marks the second splitting stage.}
\label{fig:sine_inequality_num}
\end{figure}
\FloatBarrier

\begin{remark}[Dimension-preserving readout]
\label{rem:dimreadout}
Theorem~\ref{thm:monotone} concerns the entanglement carried by the spatial
branches before they are closed. To retain the same entanglement after
recombination, the internal system must keep enough information to distinguish
those branches. After $n$ splitting stages there are $2^n$ branches, so it is
sufficient that
\[
\dim\mathcal H_X^{\mathrm{int}}\geq 2^n,
\qquad X\in\{A,B\}.
\]
Let $|\chi_i\rangle_X$ denote the internal state associated with branch
$|x_i\rangle_X$ before recombination, and let
$\{|s_i\rangle_X\}_{i=0}^{2^n-1}$ be mutually orthogonal states of the same
internal Hilbert space. Recombination may then act on the occupied subspace as
the local isometry
\[
|\chi_i\rangle_X|x_i\rangle_X
\longmapsto
|s_i\rangle_X|C\rangle_X,
\qquad i=0,\ldots,2^n-1.
\]
No branch labels are merged, so negativity is preserved exactly,
\[
\mathcal N(\rho_n^{\mathrm{int}})=\mathcal N(\rho_n),
\]
and the bounds of Theorem~\ref{thm:monotone} apply directly
to the recombined internal state.

The split--store--reuse construction of Section~\ref{sec:forward_split}
provides these states naturally: the earlier branch bits are stored in the
memory register and the final bit remains in the active two-level system. In
that case the states in Eq.~\eqref{eq:branch_register_state} are already
orthogonal, so one may take $|s_i\rangle_X=|\chi_i^{(n)}\rangle_X$. In
diamond, the NV electronic spin can serve as the reusable active system while
the host nitrogen or nearby $^{13}\mathrm C$ nuclear spins provide memory;
electron--nuclear state transfer and multi-spin control have been demonstrated
experimentally
\cite{fuchs2011memory,taminiau2014register,bradley2019register,holzgrafe2019nanodiamond}.
For $n=2$, one nuclear-memory qubit together with the active electronic qubit
provides the four orthogonal states required to retain the complete branch
label.

The following section considers the restricted case in which the same
active two-level system is reused without an auxiliary memory register. For
$n>1$ only two internal states are then available at readout, so the
positional result is no longer retained exactly.
\end{remark}

\section{Entanglement with a Two-Level Internal Readout}
\label{sec:twolevel}

We now consider a protocol with no memory register: each mass uses
the same two-level system for every split and for the final readout.
In the basis $\{|0\rangle,|1\rangle\}$, the rotation before each
split is
\begin{equation}
\label{eq:twolevel_rotation}
U=\frac{1}{\sqrt{2}}
\begin{pmatrix}1&1\\ e^{i\phi}&-e^{i\phi}\end{pmatrix}.
\end{equation}
Even and odd child branches carry $|0\rangle$ and $|1\rangle$,
respectively. A parent branch $m$ therefore enters the next
rotation in $|m\bmod 2\rangle$. The splitting phase in
Eq.~\eqref{eq:split_action} is consequently
\begin{equation}
\label{eq:twolevel_theta}
\theta_{k,m}=\phi+\pi m.
\end{equation}

\subsection{Recombination}
\label{sec:recombination}

For $n>1$, two internal states cannot retain a distinct label for
all $2^n$ spatial branches. We consider recombination in reverse
splitting order, selecting one interferometric output at each step:
\begin{equation}
\label{eq:recomb_isometry}
R_k^{(X)}|2m\rangle_X=\frac{|m\rangle_X}{\sqrt{2}},\qquad
R_k^{(X)}|2m+1\rangle_X=\frac{e^{-i\phi}|m\rangle_X}{\sqrt{2}}.
\end{equation}
The two paths interfere at one selected output. The map is therefore
not an isometry on the full input space; a second output completes
the physical recombination. Related motional operations are familiar
from Raman and Bragg interferometry
\cite{kasevich1991,cronin2009,kovachy2015}.

We use the fast-recombination approximation, neglecting the
additional gravitational phases and decoherence acquired during
closing pulses. Both protocols are evaluated at the same closure
time $t_{\mathrm{rec}}\ge\tau_n^{(2)}$. For a protocol with $m$
splits, write the binary digits of branch $i$ as
\begin{equation}
\label{eq:s4_branch_bit}
b_r(i)=\left\lfloor\frac{i}{2^r}\right\rfloor\bmod2,
\qquad r=0,\ldots,m-1,
\end{equation}
so the phases from the reverse steps combine into
\begin{equation}
\label{eq:s4_Omega}
\Omega_m(i)=\exp\!\left[
-i\phi\sum_{r=0}^{m-1}b_r(i)\right].
\end{equation}
For either $m=n$ or $m=n-1$, the selected two-level output is
\begin{equation}
\label{eq:s4_twolevel_sum}
\begin{split}
(\rho_m^{(2)})_{ab,cd}
=\frac{1}{4^m}\!
\sum_{\substack{i,j,k,l\\
i\bmod2=a,\;j\bmod2=b\\
k\bmod2=c,\;l\bmod2=d}}
&\Omega_m(i)\Omega_m(j)\\
&\times\overline{\Omega_m(k)\Omega_m(l)}
(\rho_m(t_{\mathrm{rec}}))_{ij,kl}.
\end{split}
\end{equation}
Here $a,b,c,d\in\{0,1\}$. The output $\rho_m^{(2)}$ is
unnormalised, with success probability
$p_m=\operatorname{Tr}\rho_m^{(2)}$ and, for $p_m>0$,
conditioned state $\widehat\rho_m^{(2)}=\rho_m^{(2)}/p_m$.
Recombination brings together coherences from different parity
sectors. Theorem~\ref{thm:monotone} therefore does not directly
compare the entanglement of these two readouts.

To evaluate Eq.~\eqref{eq:s4_twolevel_sum}, we express the child
matrix elements in terms of the preceding state. Write the last
branch bits as $a,b,c,d$, with parent indices $p,q,r,s$.
The final split contributes a factor $1/4$ and the local phase
$e^{i(a\theta_{n,p}+b\theta_{n,q}-c\theta_{n,r}-d\theta_{n,s})}$
to $(\rho_n)_{2p+a,\,2q+b;\,2r+c,\,2s+d}$.

For the gravitational interaction, the ket distance changes by
$(b-a)x_n(u)$, giving the extra phase $F_{b-a}(p,q)$;
the bra similarly contributes $F_{d-c}(r,s)$. Here
Eq.~\eqref{eq:phase_pair_exact} also includes the zero shift,
$F_0=0$.

The final split can also change the distance between branches
of the same mass, and hence their decoherence. The additional
decoherence exponent is
\begin{equation}
\label{eq:s4_extra_decoherence}
\begin{split}
D_n^{ab,cd}(p,q,r,s)=
\int_{\tau_n^{(1)}}^{t_{\mathrm{rec}}}\!du\,
\big[&
\Gamma_A(x_p-x_r+(a-c)x_n)
-\Gamma_A(x_p-x_r)\\
&+\Gamma_B(x_q-x_s+(b-d)x_n)
-\Gamma_B(x_q-x_s)\big],
\end{split}
\end{equation}
with all positions evaluated at $u$. This is the child's
decoherence exponent minus that of the parent. It vanishes when
$a=c$ and $b=d$, as expected within each parity sector.
Combining the three contributions gives
\begin{equation}
\label{eq:s4_rho_entries}
\begin{split}
&(\rho_n)_{2p+a,\,2q+b;\,2r+c,\,2s+d}\\
&\quad=\frac14
 e^{i(a\theta_{n,p}+b\theta_{n,q}
 -c\theta_{n,r}-d\theta_{n,s})}\\
&\qquad\quad\times
 e^{-i[F_{b-a}(p,q)-F_{d-c}(r,s)]}
 e^{-D_n^{ab,cd}(p,q,r,s)}
 (\rho_{n-1})_{pq,rs}.
\end{split}
\end{equation}
This expression includes the coherences between parity sectors,
which are combined by the readout.

\subsection{Exact Criterion for Readout Entanglement}
\label{subsec:twolevel_schur}

We now ask under what conditions an additional splitting stage
increases the negativity of the conditioned two-level readout.
The following argument holds for any number of splitting stages,
provided the final state contains two qubits.

Let $M:=\rho_n^{(2)}$ be the unnormalised output, and suppose
$p_n,p_{m}>0$. The selected recombination acts on the positive
branch-space state as $M=L\rho_nL^\dagger$, where $L$ is a local
success operator. Thus $M$ remains positive semidefinite, even
though its trace need not equal one.

In the two-qubit basis, write
\begin{equation}
\label{eq:s4_block_decomposition}
M=\begin{pmatrix}
A&X\\
X^\dagger&B
\end{pmatrix},
\qquad
M^{T_A}=\begin{pmatrix}
A&X^\dagger\\
X&B
\end{pmatrix},
\end{equation}
where $A$, $B$ and $X$ are $2\times2$ matrices. The diagonal
blocks $A,B$ are principal submatrices of $M$ and hence
$A,B\succeq0$.

For a positive two-qubit operator, the partial transpose has at
most one negative eigenvalue~\cite{rana2013}. The negativity of
the conditioned output is therefore determined by the smallest
eigenvalue of $M^{T_A}$. In particular,
$\mathcal N(\widehat\rho_n^{(2)})>
\mathcal N(\widehat\rho_{m}^{(2)})$
precisely when $M^{T_A}+qI_4$ has a negative eigenvalue, where
\begin{equation}
\label{eq:s4_schur_shift}
q=p_n\mathcal N(\widehat\rho_{m}^{(2)}).
\end{equation}

Suppose first that the preceding readout is entangled, so $q>0$.
Since $A\succeq0$, the matrix $A+qI_2$ is positive definite
and may be inverted. Completing the square for a vector with
components $u,v\in\mathbb C^2$ gives
\begin{equation}
\label{eq:s4_schur_square}
\begin{split}
&\begin{pmatrix}u\\v\end{pmatrix}^{\!\dagger}
(M^{T_A}+qI_4)
\begin{pmatrix}u\\v\end{pmatrix}\\
&=\big(u+(A+qI_2)^{-1}X^\dagger v\big)^\dagger
(A+qI_2)
\big(u+(A+qI_2)^{-1}X^\dagger v\big)\\
&\quad+v^\dagger
\big[B+qI_2-X(A+qI_2)^{-1}X^\dagger\big]v.
\end{split}
\end{equation}
The first term is nonnegative and can be made zero by choosing
$u=-(A+qI_2)^{-1}X^\dagger v$. Hence the shifted partial
transpose has a negative eigenvalue if and only if the following
$2\times2$ Schur complement has one:
\begin{equation}
\label{eq:s4_schur_complement}
K(q)=B+qI_2-X(A+qI_2)^{-1}X^\dagger.
\end{equation}
Since the partial transpose has at most one negative eigenvalue,
all its other eigenvalues become strictly positive upon adding
$qI_4$ for $q>0$. The Schur complement has the same number of
negative eigenvalues as the shifted matrix, so a negative
eigenvalue of $K(q)$ is equivalent to a negative determinant.
We thus obtain the exact condition
\begin{equation}
\label{eq:s4_n2_readout_condition}
\mathcal N(\widehat\rho_n^{(2)})>
\mathcal N(\widehat\rho_{m}^{(2)})
\quad\Longleftrightarrow\quad
\det K(q)<0,
\qquad q>0.
\end{equation}
This requires only the inverse and determinant of $2\times2$
matrices, rather than the spectrum of the full readout.

If the $m$ readout is separable, then $q=0$ and the
question is whether the new output is entangled. The block $A$
need not be invertible in that case. For any $z>0$, however,
$A+zI_2$ is invertible and $K(z)$ is defined by
Eq.~\eqref{eq:s4_schur_complement} with $q$ replaced by $z$.
The new output is entangled if and only if
$\det K(z)<0$ for all sufficiently small $z>0$.
Indeed, any negative eigenvalue of $M^{T_A}$ remains negative
under a sufficiently small positive shift, while a positive
partial transpose remains positive under every such shift.
This covers the case of a separable preceding readout.

\section{Numerical Illustration}
\label{sec:numerical}

We compare the single- and double-splitting protocols using the full
branch-space evolution in Eq.~\eqref{eq:master}. We first
calculate the negativity of the unrecombined states, then apply the
restricted no-memory two-level readout of Section~\ref{sec:twolevel}
at times $t_{\mathrm{rec}}\ge\tau_2^{(2)}$.

\medskip
\noindent\textbf{Parameters.}
We take two identical diamond spheres of mass
$m=1.5\times10^{-14}\,\mathrm{kg}$,
of the same order as the original proposal~\cite{bose2017}, with
\begin{equation*}
\kappa=G_{\mathrm N}m^2
=1.502\times10^{-38}\,\mathrm{J\,m}.
\end{equation*}
The centre separation is $d_0=500\,\mu\mathrm m$ and the two peak
path separations are $185$ and $135\,\mu\mathrm m$. These are chosen
values, guided by the need for sizeable gravitational phases while
keeping the closest branches far enough apart to limit the
Casimir--Polder interaction~\cite{bose2017,vandekamp2020}.

We use a magnetic gradient $\mathcal G=10^5\,\mathrm{T\,m^{-1}}$
and an effective moment $\mu_{\mathrm{eff}}=2\mu_{\mathrm B}$,
corresponding to the NV electronic-spin states. Margalit et al. propose generating magnetic gradients of order
$8.7\times10^4\,\mathrm{T\,m^{-1}}$ using current-carrying
wires on an atom chip~\cite{margalit2021}. Maintaining such
gradients over the trajectories and pulse durations considered
here remains an experimental challenge. Alternative field
geometries for producing large spatial superpositions of
mesoscopic masses have also been explored
\cite{marshman2022,scala2013}.
For a branch-preserving implementation, earlier spin labels could
be stored locally while the electronic spin is reused
\cite{fuchs2011memory,taminiau2014register,bradley2019register}.

\begin{figure}[p]
\centering
\includegraphics[width=\linewidth]{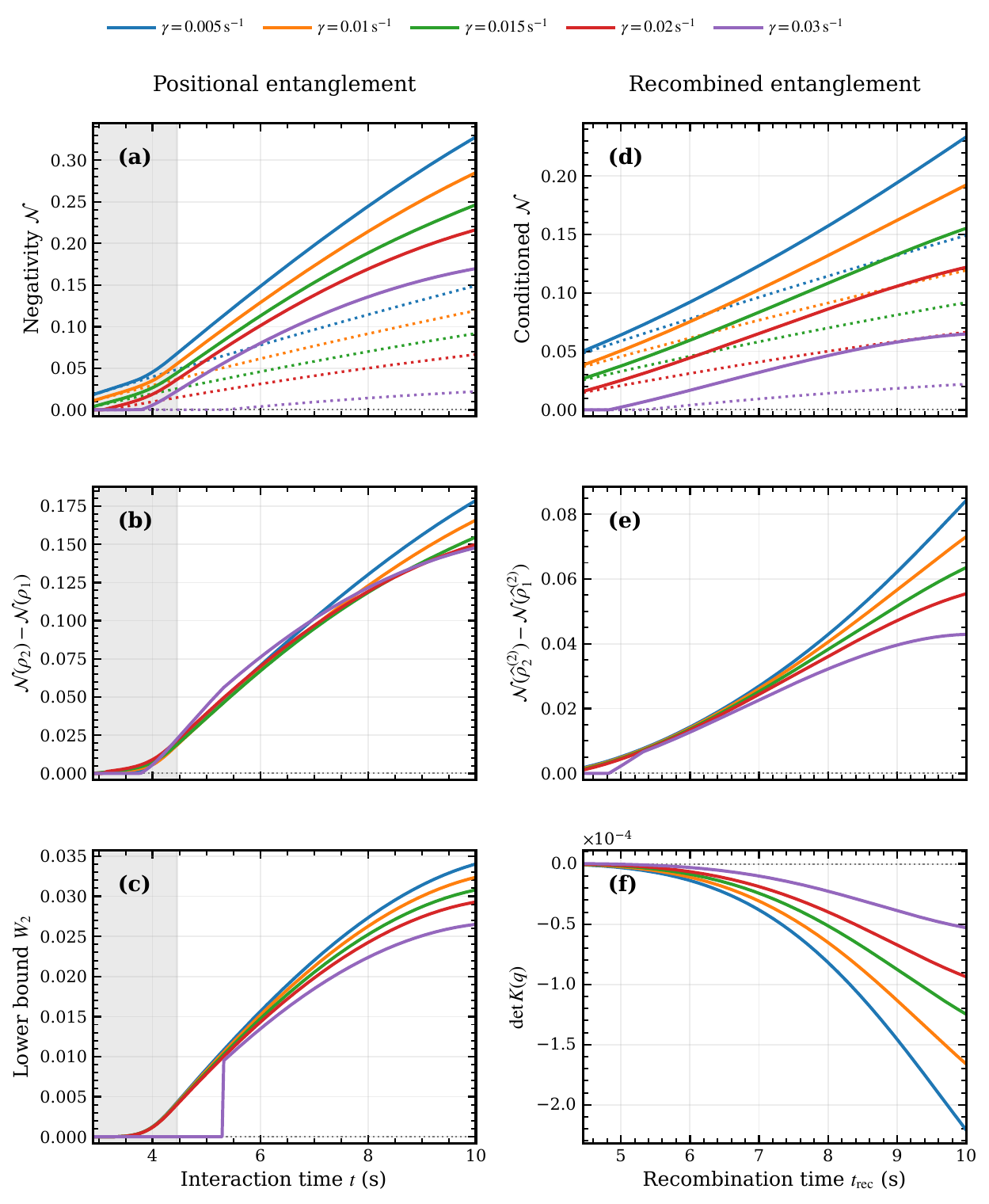}
\caption{Numerical comparison of the single- and double-splitting protocols for the parameter set in Table~\ref{tab:params}. Left column: positional entanglement before recombination. Right column: conditioned entanglement after the restricted two-level readout. Dotted curves denote one splitting stage and solid curves denote two. Panels \textbf{(a)} and \textbf{(d)} show the negativities, \textbf{(b)} and \textbf{(e)} show the corresponding gains, \textbf{(c)} shows the lower bound $W_2$, and \textbf{(f)} shows the Schur determinant $\det K(q)$. Shaded intervals mark the second split.}
\label{fig:numerical_comparison}
\end{figure}

The drift times are $T_{d,1}=2.0\,\mathrm s$ and
$T_{d,2}=0.5\,\mathrm s$, with an inter-stage delay
$\delta=0.25\,\mathrm s$. From Eq.~\eqref{eq:x_final}, these
choices give
\begin{align*}
T_{\alpha,1}&\simeq0.3221\,\mathrm s,
&T_{\alpha,2}&\simeq0.5300\,\mathrm s,\\
\tau_1^{(2)}&\simeq2.6443\,\mathrm s,
&\tau_2^{(1)}&\simeq2.8943\,\mathrm s,\\
\tau_2^{(2)}&\simeq4.4543\,\mathrm s.
\end{align*}
We evaluate the evolution up to $t=10\,\mathrm s$, allowing both
protocols to accumulate appreciable gravitational phases.
In Eq.~\eqref{eq:gamma}, we use
$a_{\mathrm{loc}}=1\,\mathrm{nm}$ and
$\gamma=0.005,0.010,0.015,0.020,0.030\,\mathrm{s^{-1}}$.
These are assumed phenomenological rates, chosen to examine the
regime studied in related decoherence analyses
\cite{schut2022,tilly2021,marshman2022}. The parameters are summarised in
Table~\ref{tab:params} with results shown in Figure~\ref{fig:numerical_comparison}.

\begin{table}[htbp]
\centering
\caption{Parameters of the numerical comparison. Literature
references motivate the physical scales; the exact geometry,
pulse timings and decoherence rates are modelling choices.}
\label{tab:params}
\small
\begin{tabularx}{\linewidth}{@{}l l l >{\raggedright\arraybackslash}X@{}}
\toprule
Parameter & Symbol & Value & Basis \\
\midrule
Mass & $m$ & $1.5\times10^{-14}\,\mathrm{kg}$ &
Proposal scale~\cite{bose2017} \\
Gravitational coupling & $\kappa$ &
$1.502\times10^{-38}\,\mathrm{J\,m}$ & $G_{\mathrm N}m^2$ \\
Centre separation & $d_0$ & $500\,\mu\mathrm m$ &
Chosen; cf.~\cite{bose2017,vandekamp2020} \\
Magnetic gradient & $\mathcal G$ & $10^5\,\mathrm{T\,m^{-1}}$ &
Proposed nearby-wire scale~\cite{margalit2021} \\
Effective moment & $\mu_{\mathrm{eff}}$ & $2\mu_{\mathrm B}$ &
NV-spin splitting~\cite{bose2017,margalit2021} \\
\midrule
First peak separation & $x_1^{\mathrm{peak}}$ & $185\,\mu\mathrm m$ &
Chosen; cf.~\cite{bose2017,marshman2022} \\
First drift & $T_{d,1}$ & $2.0\,\mathrm s$ &
Chosen pulse duration \\
First acceleration & $T_{\alpha,1}$ & $0.3221\,\mathrm s$ &
From Eq.~\eqref{eq:x_final} \\
First split ends & $\tau_1^{(2)}$ & $2.6443\,\mathrm s$ &
Calculated \\
Inter-stage delay & $\delta$ & $0.25\,\mathrm s$ &
Chosen local-control interval \\
Second split begins & $\tau_2^{(1)}$ & $2.8943\,\mathrm s$ &
Calculated \\
Second peak separation & $x_2^{\mathrm{peak}}$ & $135\,\mu\mathrm m$ &
Chosen; cf.~\cite{bose2017,marshman2022} \\
Second drift & $T_{d,2}$ & $0.5\,\mathrm s$ &
Chosen pulse duration \\
Second acceleration & $T_{\alpha,2}$ & $0.5300\,\mathrm s$ &
From Eq.~\eqref{eq:x_final} \\
Second split ends & $\tau_2^{(2)}$ & $4.4543\,\mathrm s$ &
Calculated \\
\midrule
Localisation length & $a_{\mathrm{loc}}$ & $1\,\mathrm{nm}$ &
Assumed in Eq.~\eqref{eq:gamma} \\
Decoherence rate & $\gamma$ &
$0.005$--$0.030\,\mathrm{s^{-1}}$ &
Model sweep; cf.~\cite{schut2022,tilly2021} \\
Evaluation time & $t_{\max}$ & $10\,\mathrm s$ &
Chosen interaction window \\
\bottomrule
\end{tabularx}
\end{table}

\FloatBarrier

\FloatBarrier

\medskip

\noindent\textbf{Results.}
Figure~\ref{fig:numerical_comparison} shows that two splits produce greater
branch-space negativity than one split at every decoherence rate
considered. Panel~(c) also shows a positive lower bound $W_2$ once
the parent state has a negative partial-transpose eigenvalue, so
Theorem~\ref{thm:monotone} certifies part of the improvement.
At $t=10\,\mathrm s$ and the highest rate
$\gamma=0.030\,\mathrm{s^{-1}}$, the single- and double-split
negativities are $0.02199$ and $0.16978$, respectively, while
$W_2\simeq0.02651$. The parent therefore remains entangled even
in the most strongly decohering case shown.

The second split improves the readout negativity for all
five rates at $t_{\mathrm{rec}}=10\,\mathrm s$.
For $\gamma=0.030\,\mathrm{s^{-1}}$, the single- and double-split
conditioned negativities are $0.02199$ and $0.06489$, respectively.
The corresponding success probability is $p_2\simeq0.06081$, and
the Schur determinant is $\det K(q)\simeq-5.29\times10^{-5}$.
Panel~(f) shows the negative determinant that certifies the strict
readout improvement, in agreement with
Eq.~\eqref{eq:s4_n2_readout_condition}. At
$\gamma=0.015\,\mathrm{s^{-1}}$, the conditioned negativities are
$0.09178$ and $0.15531$, with $p_2\simeq0.06022$ and
$\det K(q)\simeq-1.25\times10^{-4}$. This success probability
must be taken into account when comparing experimentally observed
yields rather than conditioned entanglement only.

The model includes the gravitational phases and the local
decoherence law in Section~\ref{sec:time_evolution}.
It does not include imperfect spin or memory control, branch
misalignment during recombination, fluctuating magnetic gradients,
acceleration and tilt noise, or residual electromagnetic forces;
these can limit an experimental implementation
\cite{toros2021,moorthy2026random,moorthy2025magnetic,
wu2025torsion,fragolino2024dipole,schut2023screening}.
These are omitted here to isolate what changes when an additional
coherent splitting stage is introduced. 
At the minimum branch separation of $180\,\mu\mathrm m$, the
retarded Casimir--Polder potential is estimated to be about
$4.4\%$ of the Newtonian potential; this contribution is
neglected in our simulations~\cite{vandekamp2020,schut2024micrometer}.

\section*{Conclusion}
Repeated splitting provides more paths through which two masses can
acquire gravitationally induced phases. The parity structure makes
it possible to compare consecutive stages without diagonalising
the larger state. After local parity dephasing, two sectors reproduce
the preceding state, while the other two differ only by known
additional gravitational phases. Removing the locally generated
parts gives an exact lower bound on the change in negativity from
the negative eigenspace of the preceding state's partial transpose.
The same block structure also shows that an additional split cannot
remove all previously generated entanglement.

For two splitting stages, the positional bound reduces to an
explicit inequality involving three gravitational phases. Our
numerical example satisfies this condition and shows greater
unrecombined negativity across the decoherence rates considered.
For a two-level readout without memory, we express the recombined
state directly through the interaction and decoherence integrals.
An exact test based on a $2\times2$ Schur complement then determines
whether an additional split improves the conditioned negativity without
requiring the spectrum of the full partial transpose. The numerical
example satisfies this test at all decoherence rates considered. The successful
readout is probabilistic and need not retain all the positional
entanglement. A register that preserves the full branch label, by
contrast, retains it exactly.

The two-split condition is especially simple because the preceding
state has only two branches per mass. One natural next step is to seek
similarly explicit conditions for higher numbers of splits. The general
one-step bound already applies, but evaluating it then requires the
negative eigenspace of a larger parent state. Another direction is to
optimise the number of splits, their separations, pulse durations and
delays together, rather than selecting a geometry in advance. Such an
optimisation should compare the entanglement in the spatial branches
with that retained after recombination, including the probability of
successful readout. It would also be interesting to combine repeated
splitting with qudit strategies proposed to improve resilience to
decoherence~\cite{tilly2021}, allowing the positions, amplitudes and
phases of the additional paths to be chosen together. These extensions
would help establish when extra splitting is most useful and how much
of its advantage can be accessed experimentally. Moreover, one could also optimise experiments to offer the most discrimination between competing theories, rather than optimising entanglement generation in a single framework. 

\section*{Acknowledgments} S.B’s research is funded by the Gordon and
Betty Moore Foundation through Grant GBMF12328, DOI
10.37807/GBMF12328, and the Alfred P. Sloan Foundation
under Grant No. G-2023-21130. This work was made possible through the support of the WOST,
WithOut SpaceTime project (https://withoutspacetime.org),
supported by Grant ID \char"0023 63683 from the John Templeton Foundation (JTF). 

\begin{multicols}{2}[\section*{References}]
\printbibliography[heading=none]
\end{multicols}

\end{document}